\documentclass[%
 reprint,
 amsmath,
 amssymb,
 prl,
]{revtex4-1}
\usepackage{graphicx}% Include figure files
\usepackage{dcolumn}% Align table columns on decimal point
\usepackage{bm}% bold math
\usepackage{lipsum} 

\begin{document}

\title{Generation of forerunner electron beam during interaction of ion beam pulse with plasma} %Title of paper

\author{Kentaro Hara}
\email{khara@tamu.edu}
%\affiliation{Department of Aerospace Engineering, Texas A\&M University, College Station, TX, 77843-3141}
\author{Igor D. Kaganovich}
\email{ikaganov@pppl.gov}
\affiliation{Princeton Plasma Physics Laboratory, Princeton University, Princeton, NJ 08543}
\author{Edward A. Startsev}
\affiliation{Princeton Plasma Physics Laboratory, Princeton University, Princeton, NJ 08543}

\date{\today} 
 
\begin{abstract}
The long-time evolution of the two-stream instability of a cold ion beam pulse propagating though the background plasma is investigated using a large-scale one-dimensional electrostatic kinetic simulation. The three stages of the instability are identified and investigated in detail. After the initial linear growth and saturation by the electron trapping, a portion of the initially trapped electrons becomes detrapped and moves ahead of the ion beam pulse forming a {forerunner} electron beam, which causes a secondary two-stream instability that preheats the upstream plasma electrons. Consequently, the self-consistent nonlinear-driven turbulent state is set up at the head of the ion beam pulse with the saturated plasma wave sustained by the influx of the cold electrons from the upstream of the beam that lasts until the final stage when the beam ions become trapped by the plasma wave. The beam ion trapping leads to the nonlinear heating of the beam ions that eventually extinguishes the instability. 
\end{abstract}
\maketitle

%%%%%%%%%%%%%%%%%%%%%%%%%%%%%%%%%%%%%%%%%%%%%%%%%%%%%%%%%%%%%%%%%%%%%%%%
%\section{Introduction} 
%%%%%%%%%%%%%%%%%%%%%%%%%%%%%%%%%%%%%%%%%%%%%%%%%%%%%%%%%%%%%%%%%%%%%%%%
\textit{Introduction.} The two-stream instability plays an important role in fusion~\cite{roth01, deutsch04, friedman09, olson14}, astrophysics~\cite{cargill88, dieckmann00}, double layer formation~\cite{iizuka79, sato80}, and thrusters~\cite{campanell12, jorns14}. In particular, nonrelativistic ion beams can be used for heavy ion fusion and  warm-dense matter experiments~\cite{seidl09, seidl15, busold15}. Neutralization of the ion beam is particularly important for the beam quality as the space charge may defocus the beam~\cite{kaganovich04, kaganovich07, kaganovich08}, which has been studied for under-dense~\cite{kaganovich07} and tenuous~\cite{berdanier15} plasmas. Longitudinal~\cite{roy05,startsev14} and transverse compression~\cite{sudan76, dorf09, mitrani14, tokluoglu15} have also been investigated to increase the ion beam density. 

A neutralized ion beam triggers {an electrostatic} two-stream instability between beam ions and plasma electrons; the instability saturates due to wave-particle trapping of either beam ions or plasma electrons~\cite{startsev14}. Some fraction of the wave-trapped electrons  become detrapped and streams ahead of the neutralized ion beam pulse. {This} results in generation of a beam of accelerated electrons propagating though the background plasma{, which we call forerunner electrons}. As a consequence, a secondary two-stream instability is developed between the accelerated and background electrons. 

Because two-stream instability can strongly affect ion beam ballistic propagation in the background plasma, it is important to investigate the long-time evolution of the secondary instability. The saturation of the initial two-stream instability by wave trapping has been investigated in previous studies~\cite{startsev14, tokluoglu15} where a small computational domain around the beam pulse was used to perform two-dimensional simulations. The effect of the streaming electrons ahead of the beam pulse {(i.e., electron acceleration and wave decay processes)} was not thoroughly investigated. Recent simulations show that large spatial domain and long temporal simulations are essential to investigating the long-time dynamics of the beam-plasma interactions~\cite{brunner14, park15}. The focus of this Letter is to study the later phase of the two-stream instability -- (i) how the electrons become detrapped from the wave and accelerate ahead of the ion beam pulse and (ii) how they affect long-time evolution of the initial two-stream instability. Therefore,  we report the results of a large-scale, one-dimensional electrostatic kinetic simulation of the interaction between the ion beam pulse and background plasma.

\textit{Kinetic simulation.} Electrostatic kinetic simulations are performed in the frame of the ion beam. 
A standard particle-in-cell (PIC) simulation~\cite{birdsall} is used for the ion beam pulse, the background ions, and the background electrons. The cell size is $\Delta x = L/N_x$, where  $L=15$ m is the domain length and  $N_x = 3 \times 10^4$ is the number of cells. Li$^+$ is assumed for the ions. The electron temperature is 0.4 eV; the ion temperature is 0.3 eV; and the ion beam temperature is 0 eV. The ion beam density profile is assumed to be a Gaussian pulse with a duration of 20~ns. The plasma density is $n_p = 5.5 \times 10^{16}$ m$^{-3}$, the ion beam density is  $n_b =2 \times 10^{15}$ m$^{-3}$, and the ion beam velocity is chosen to be $v_b = c/30$, where $c$ is the speed of light; the beam and plasma parameters are similar to the neutralized drift compression experiment (NDCX) parameters~\cite{tokluoglu15}. The boundary conditions for the Poisson equation is  $\phi=0$ and $\partial_x\phi = 0$  at the boundary in front of the beam. The presented results are checked for convergence using small grid sizes (0.1 mm) and a large number of computational particles (3000 particles per cell), as well as with a separate Vlasov simulation solver~\cite{hara15pop, hara17} with comparable grid sizes in phase space.

\begin{figure*} [t!]
\includegraphics[width=0.9 \linewidth]{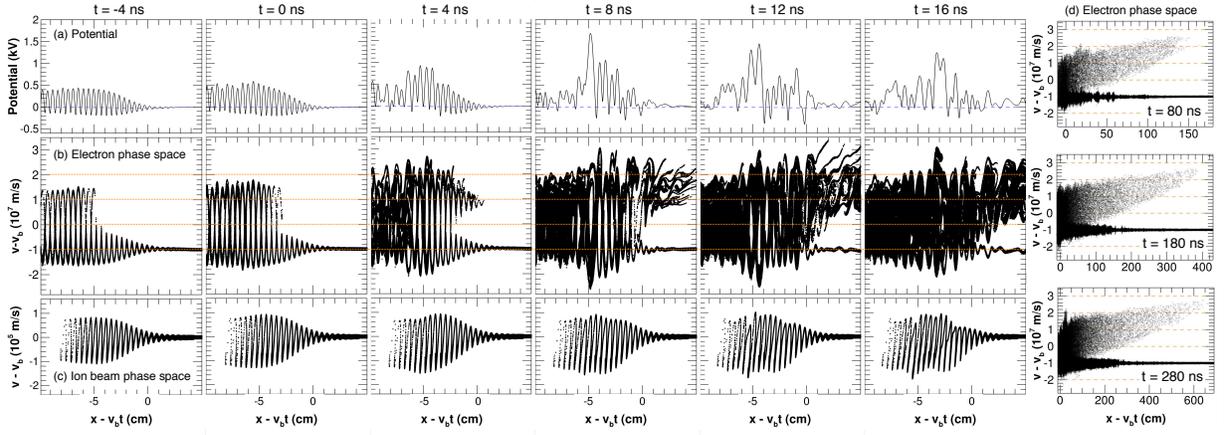} 
\caption{Overview of the electron acceleration due to the two-stream instability caused by the neutralized ion beam at different times indicated in the top legends, a small 15cm long window out of a 15m  long computational domain is shown. Shown are (a) the potential, (b) the phase space of electrons, and (c) the phase space of the ion beam in the beam frame, $v-v_b, x-v_bt$. {The electron phase space at three different times are shown in (d).} $t=0$ is chosen to be the time when forward moving electrons{, namely, the forerunner electrons,} are generated after the saturation of the initial instability, which is approximately 200~ns after the injection of the ion beam pulse into the plasma. }
\label{fig:200ns}
\end{figure*}

\textit{{Multiple} stages of the two-stream instability.} After the beam is injected into a plasma, the two-stream instability develops and  saturates nonlinearly. Figure~\ref{fig:200ns} shows potential,  electron phase space, and ion beam phase space in the beam frame. Several stages of evolution of the two-stream instability between the beam ions and plasma electrons can be observed in Fig.~\ref{fig:200ns}.  We focus on nonlinear stages of instability 200~ns after injection into a plasma, when we initialize time $t=0$ presented in all figures. At $t\leq 0$~ns, the potential modulations are relatively small and confined to the beam pulse region ($|x-v_b t|<10$~cm). The growth rate of the ion-beam induced two-stream instability~\cite{buneman59} is $\gamma/\omega_{pe} \approx (\sqrt{3}/2 ) \sqrt[3]{n_b/n_p \cdot m_e/m_{i,b}}$, where $\omega_{pe}=\sqrt{4 \pi e^2 n_p/ m_e}$ ($e$, $m_e$, and $m_{i,b}$ being the electric charge, electron mass, and beam ion mass). The phase velocity of plasma wave and the wavelength of the modulation agree with theoretical predictions~\cite{buneman59}: $v_\phi - v_b = - (\gamma/\sqrt{3})/k = -5.6 \times 10^4$ and $L = 2\pi v_b /  \omega_{pe} \approx 4.8$ mm. The potential amplitude grows until saturation due to electron trapping~\cite{tokluoglu17}. The potential in the plasma wave becomes very asymmetric and reaches about 1.6 kV at maximum at $t=8$~ns. Electron trapping can be clearly observed in plasma electron phase plots shown in Fig.~\ref{fig:200ns}(b), where the electron velocity modulation reaches levels of ion beam velocity, $v_b$.  Around $t\sim0$~ns, the wave breaking causes the potential structure to become incoherent and nonstationary in the beam frame. At this time, electrons become detrapped, escape from the potential wells  in the plasma wave, and are accelerated ahead of the beam pulse forming a forerunner electron beam. After the electron acceleration occurs, the potential amplitude gradually decreases.  Additionally, the newly generated electron stream causes a secondary two-stream instability between the streaming electrons and the background plasma electrons [see Fig.~\ref{fig:200ns} $x-v_b t > 2$~cm and $t\geq 8$~ns]. This instability growth rate is much faster than the initial ion-beam instability because the secondary instability is between two electron populations: $\gamma_s/ \omega_{pe} \propto \sqrt[3]{n_s / n_p}$, where  $n_s$ is the density of the forerunner electron beam. {The growth rate occurs on the ns-scale} although $n_s/n_b \approx O(10^{-2})$.  {Self-similar evolution of forerunner electron beams is shown in Fig.~\ref{fig:200ns}(d). It can be seen that electrons are constantly being accelerated near the ion beam region but will experience heating due to the secondary two-stream instability.}   

\begin{figure} [b!]
\includegraphics[width=0.81 \linewidth]{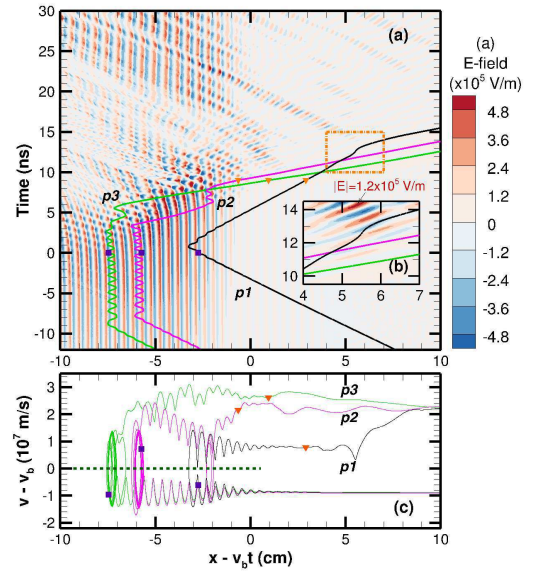}
\caption{Spatio-temporal evolution of the electric field and the trajectories of three test particles. Particle 1 (p1) is one of the first electrons that are reflected in front of the ion beam pulse. Particles 2 (p2) and 3 (p3) experience trapping and detrapping before being accelerated in front of the ion beam pulse.  Blue and orange symbols are the location at $t=0$~ns and $t=9$~ns, respectively. {Insert (b) displays zoom in of the orange box in Fig.~\ref{fig:4}(a), showing additional acceleration of  p1 by the plasma wave generated by the forerunner beam.}}
\label{fig:4}
\end{figure}

%%%%%%%%%%%%%%%%%%%%%%%%%%%%%%%%%%%%%%%%%%%%%%%%%%%%%%%%%%%%%%%%%%%%%%%%

\begin{figure} [t!]
\includegraphics[width=0.82 \linewidth]{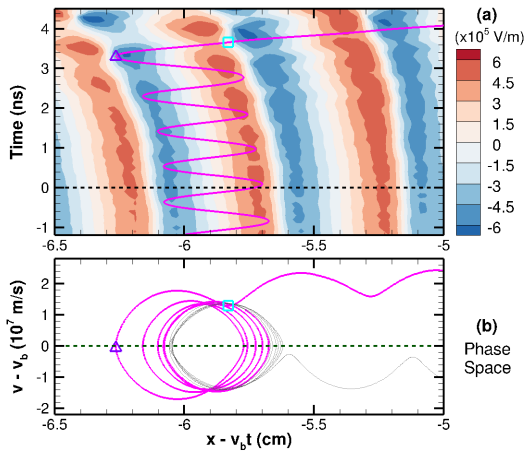}
\caption{ Zoom-up view for p2 from Fig.~\ref{fig:4}. The pink solid line is the trajectory for $t\geq -1.2$~ns. Purple and light blue symbols show the electron location at $t=3.3$~ns and $t=3.8$~ns (when the electron detrapping occurs), respectively. }
\label{fig:5}
\end{figure}

\textit{Electron acceleration due to two-stream instability.} As shown in Fig. ~\ref{fig:4}, particles 2 (p2) and 3 (p3) are initially trapped by the plasma wave at the tail of the ion beam pulse ($x-v_b t<-5$~cm) for a few cycles. Because of the wave breaking at $t>5$~ns, the electric field in the plasma wave becomes incoherent and accelerating and decelerating cycles of the electric field become asymmetric (see Fig.~\ref{fig:5}), which causes the particles to escape trapping in the wave and accelerate to move faster than the ion beam. The resulting velocity of accelerated particles lies in the interval $v - v_b \in[0 , 2v_b]$ in the beam frame (see Fig.~\ref{fig:200ns}), therefore {the generated forerunner beam travels faster than a mere reflection from  potential wells, $v>2v_b$ in the lab frame, as illustrated by} particle p1. The p1 trajectory is nearly symmetric around $v-v_b=0$ in the phase space, which indicates that p1 is purely reflected by the large-amplitude plasma wave. For most accelerated electrons, the energy builds up by particle trapping and detrapping in the waves. Note that p1 is further accelerated at $x-v_b t=5.5$~cm ($t=13$~ns) in such a process as shown in Fig.~\ref{fig:200ns}(b). In addition, there are particles that lose energy in this process, for example, this happen to particle 2 (p2) at time around $5$~ns as evident in Fig.~\ref{fig:4}(c). Once the detrapped particles, e.g., p2 and p3, form the forerunner beam, the electric field is modulated due to the secondary two-stream instability excited by the streaming electrons (at $t>8$~ns; see Fig.~\ref{fig:200ns}). This wave is responsible for additional acceleration of reflected particles (e.g., p1) from $v_b$ to $(1\sim 2)v_b$ in the beam frame, as shown in Figs.~\ref{fig:4}(b) and ~\ref{fig:4}(c).

Further details of electron acceleration are given in Fig.~\ref{fig:5}, which is in the zoomed-in part of Fig.~\ref{fig:4}. It can be seen in Fig.~\ref{fig:5} that the p2 electron gains energy and becomes accelerated forward  by moving into a negative electric field, $E_{\rm min}=-600$~kV/m, shown by the purple triangle symbol in Fig.~\ref{fig:5} at $x-v_b t=-6.26$~cm ($t=3.3$~ns), which is considerably enhanced compared to the previous bounce period. After being accelerated, the electrons move through the region of a smaller decelerating field,  $E_{\rm max}=300$~kV/m, shown by the light blue square symbol in Fig.~\ref{fig:5} at $x-v_b t=-5.83$~cm ($t=3.8$~ns). This field is weaker than the accelerating field; therefore the electrons become detrapped from the potential well and are being accelerated ahead of the beam pulse. In Fig.~\ref{fig:4}, coherent plasma waves are observed near the ion beam pulse  at  $t>18$~ns{, long after the generation of forerunner electron beam at $t\sim0$~ns.} These waves also experience modulation, which allows for the electron acceleration to occur even at later time and continuous generation of the forerunner electron beam.

\begin{figure} [t!]
\includegraphics[width=0.85 \linewidth]{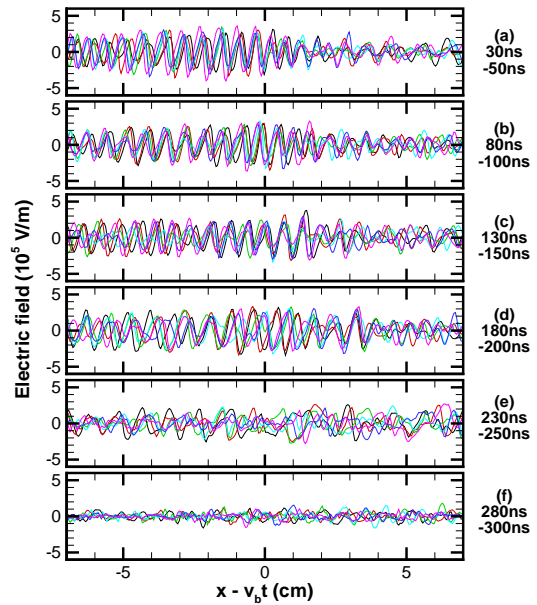}
\caption{{Long-time evolution of} plasma wave near the ion beam region at $-7$ cm $<x-v_b t<7$ cm after  {generation of forerunner electron beam}. Six color lines (4~ns apart) are overlapped in each subfigures in the order of black, red, green, light blue, blue, and pink. Coherent plasma waves are observed at $x-v_b t<3$ cm, indicating formation of stationary plasma wave at $t\leq200$~ns. The wave in front of the ion beam is more chaotic.}
\label{fig:zoomup}
\end{figure}

%%%%%%%%%%%%%%%%%%%%%%%%%%%%%%%%%%%%%%%%%%%%%%%%%%%%%%%%%%%%%%%%%%%%%%%%
\textit{Saturation and decay of the instabilities.} Figure \ref{fig:zoomup} shows the temporal and spatial structures of  the plasma wave  at $30 \leq t \leq 300$~ns.  From this figure, it is evident that the plasma wave amplitude remains relatively constant until the wave starts to decay at $t>200$~ns. This enables the high-energy ion beam to transfer its energy into the plasma electrons for a long period of time.

\begin{figure} [t!]
\includegraphics[width=0.9 \linewidth]{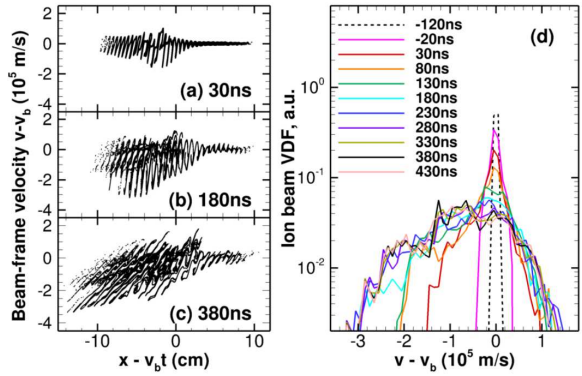}
\caption{Long-time evolution of ion beam in phase space at 30~ns (a), 180~ns (b), and 380~ns (c); and the averaged ion distribution for various time steps (d).}
\label{fig:ionbeam}
\end{figure}

Figure~\ref{fig:ionbeam} shows the temporal evolution of the phase-space of the ion beam pulse (Figs.~\ref{fig:ionbeam}(a)-(c)) and the ion velocity distribution function (IVDF) that is averaged over entire beam pulse (Fig.~\ref{fig:ionbeam} (d)).   It can be seen from Figs.~\ref{fig:ionbeam}(a)-(b) that the ions are  being trapped in the plasma wave within the first 200~ns  (the minimum ion beam velocity reaches approximately $v-v_b = -3 \times 10^5$ m/s). The ion trapping occurs because there is a coherent plasma wave that is nearly stationary in the beam frame (see Fig.~\ref{fig:zoomup}). At $t>200$~ns,  strong phase mixing leads to heating of the ion beam  (see  Fig.~\ref{fig:ionbeam}(c)). At that time the plasma waves start decaying due to weakening of the two-stream instability, { because of the thermalization of the ion beam}. The initial ion beam temperature is 0 eV and increases to approximately 1 keV at later time. Figure~\ref{fig:ionbeam}(d) shows that the mean velocity of the ion beam slows down because the ion beam energy is transferred to the electrons and plasma waves. For a sinusoidal periodic wave, the bounce frequency of the {trapped} beam ions in the plasma wave is given by $\omega_{B,i} =  k (e\phi_{\rm max}/m_{i,b})^{1/2}$, where $k$ is the wavenumber and $\phi_{\rm max}$ is the potential amplitude. The plasma potential, $e\phi_{\rm max} \propto m_e v_b^2$~\cite{kaganovich04} and $k\approx \omega_{pe}/v_b$. Therefore, the ion trapping time can be written as $ \tau_{B,i} \equiv  2\pi/\omega_{B,i}$, where $\omega_{B,i}=(4\pi e^2 n_p/m_{i,b})^{1/2}$, which is independent of the ion beam velocity. From our simulation results, $\tau_{B,i}  \approx 200$~ns, which is in good agreement with the time required for the saturation of instability as can be seen from Figs.~\ref{fig:zoomup}(e). 

\begin{figure} [t!]
\includegraphics[width=0.78 \linewidth]{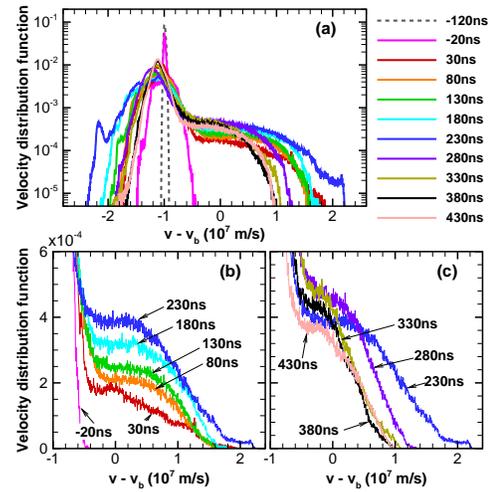}
\caption{Long-time evolution of electron VDFs averaged over space in the ion beam pulse region, i.e., $-10$~cm $<x<$ 10~cm for different times during beam propagation; (b) and (c) are zoom-in into high velocity tail region.  }
\label{fig:evdf}
\end{figure}

Figure~\ref{fig:evdf} shows the temporal evolution of the spatially averaged electron VDFs in the ion beam pulse region. The accelerated electron density increases before $t=230$~ns and decreases after $t=230$~ns, as can be seen from Figs.~\ref{fig:evdf}(b) and \ref{fig:evdf}(c) due to wave decay after $t=230$~ns. The electron trapping time is given by $\tau_{B,e} = 2 \pi / \omega_{B,e} \propto 2 \pi / \omega_{pe}$,  which is on the order of a nanosecond. {In Fig.~\ref{fig:evdf}(a), heating of the background electrons can also be observed up to $t=230$~ns, which is due to the secondary two-stream instability.} Note that a significant amount of electrons is accelerated and the position of the maximum of the VDF is shifted toward the negative velocity, so that the total current is maintained, i.e., the current of the ion beam pulse is fully neutralized by the plasma electrons in 1D case. This may be different in a multidimensional setup if the beam radius is small compared to the skin depth, because the electron acceleration can occur along the beam axis and the return current may occur outside the beam~\cite{tokluoglu17}. 

\textit{Summary.}  We performed large spatial and long temporal studies of the two-stream instability produced by an ion beam pulse propagating in the background plasma using a one-dimensional electrostatic kinetic simulation.  After the initial linear stage of the instability is terminated by the electron trapping, some of the electrons are accelerated by the strong plasma wave to  about twice the beam velocity and propagate ahead of the ion beam pulse. Hence, we call it the forerunner electron beam. Examination of the electron trajectories forming the forerunner beam shows that the acceleration mostly occurs due to the energy gain during the electrons trapping and detrapping in the nonstationary plasma wave setup after the initial saturation. The strong plasma wave driven by the influx of the cold electrons from upstream persists for the time of the order of the ion bounce period in the this nonlinear plasma wave  ($\tau_{B,i}\propto 2\pi/(4\pi e^2 n_p/m_{i,b})^{1/2}$)  and only  decays when  the beam ions become trapped  and heated by the action of the wave. During this time the continuous generation of the forerunner electron beam was observed. The forerunner electron beam can strongly preheat background plasma. The ion beam propagates distance $v_b/\tau_{B,i}$ during the time $\tau_{B,i}$. Therefore, the strong defocusing forces caused by the two stream instability~\cite{startsev14, tokluoglu17} can affect the ballistic beam propagation in plasmas only on distances shorter than $v_b/\tau_{B,i}$.

\textit{Acknowledgement.} Authors thank  fruitful discussions with  E. Tokluoglu, A. Khrabrov, and J. Carlson. This research was funded by the U.S. Department of Energy.


\begin{thebibliography}{10}%
\bibitem{roth01}%
M.~Roth, T.~E.~Cowan, M.~H.~Key, S.~P.~Hatchett,
C.~Brown, W.~Fountain, J.~Johnson, D.~M.~Pennington,
R.~A.~Snavely, S.~C.~Wilks, K.~Yasuike, H.~Ruhl, F.~Pegoraro, S.~V.~Bulanov, E.~M.~Campbell, M.~D.~Perry, and H.~Powell, Phys. Rev. Lett. {\bf 86}, 436 (2001).

\bibitem{deutsch04}%
C.~Deutsch, Laser Part. Beams. {\bf 22}, 115 (2004).

\bibitem{friedman09}
A.~Friedman, J.~Barnard, R.~Briggs, R.~Davidson,
M.~Dorf, D.~Grote, E.~Henestroza, E.~Lee, M.~Leitner,
B.~Logan, A.~Sefkow, W.~Sharp, W.~Waldron, D.~Welch,
and S.~Yu, Nucl. Instrum. Meth. A  {\bf 606}, 6 (2009).

\bibitem{olson14}
C.~L.~Olson, Nucl. Instrum. Meth. A  {\bf 733}, 86 (2014).

\bibitem{cargill88}
P.~J.~Cargill and K.~Papadopoulos, Astrophys. J  {\bf 329}, L29 (1988).

\bibitem{dieckmann00}
M.~ E.~Dieckmann, P.~Ljung, A.~Ynnerman, and K.~G.~ McClements, Phys. Plasmas {\bf 7}, 5171 (2000).

\bibitem{iizuka79}
 S.~Iizuka, K.~Saeki, N.~Sato, and Y.~Hatta, Phys. Rev.
Lett. {\bf 43}, 1404 (1979).

\bibitem{sato80}
 T.~Sato and H.~Okuda, Phys. Rev. Lett. {\bf 44}, 740 (1980).

\bibitem{campanell12}
 M.~D.~Campanell, A.~V.~Khrabrov, and I.~D.~Kaganovich,
Phys. Rev. Lett. {\bf 108}, 235001 (2012).

\bibitem{jorns14}
 B.~A.~Jorns, I.~G.~Mikellides, and D.~M.~Goebel, Phys.
Rev. E {\bf 90}, 063106 (2014).

\bibitem{seidl09}
 P.~Seidl, A.~Anders, F.~Bieniosek, J.~Barnard, J.~Calanog,
A.~Chen, R.~Cohen, J.~Coleman, M.~Dorf, E.~Gilson,
D.~Grote, J.~Jung, M.~Leitner, S.~Lidia, B.~Logan, P.~Ni,
P.~Roy, K.~V.~den~Bogert, W.~Waldron, and D.~Welch,
Nucl. Instrum. Meth. A {\bf 606}, 75 (2009).

\bibitem{seidl15}
P.~A.~Seidl, A.~Persaud, W.~L.~Waldron, J.~J.~Barnard,
R.~C.~Davidson, A.~Friedman, E.~P.~Gilson, W.~G.~Greenway, D.~P.~Grote, I.~D.~Kaganovich, S.~M.~Lidia, M.~Stettler, J.~H.~Takakuwa, and T.~Schenkel, Nucl. Instrum. Meth. A {\bf 800}, 98 (2015).

\bibitem{busold15}
 S.~Busold, D.~Schumacher, C.~Brabetz, D.~Jahn, F.~Kroll,
O.~Deppert, U.~Schramm, T.~E.~Cowan, A.~bel~Blazevic,
V.~Bagnoud, and M.~Roth, Sci. Rep {\bf 5}, 12459
(2015).

\bibitem{kaganovich04}
 I.~D.~Kaganovich, E.~A.~Startsev, and R.~C.~Davidson,
Phys. Plasmas {\bf 11}, 3546 (2004).

\bibitem{kaganovich07}
I.~D.~Kaganovich, E.~A.~Startsev, A.~B.~Sefkow, and
R.~C.~Davidson, Phys. Rev. Lett. {\bf 99}, 235002 (2007).

\bibitem{kaganovich08}
I.~D.~Kaganovich, E.~A.~Startsev, A.~B.~Sefkow, and
R.~C.~Davidson, Phys. Plasmas {\bf 15}, 103108 (2008).

\bibitem{berdanier15}
 W.~Berdanier, P.~K.~Roy, and I.~D.~Kaganovich, Phys. Plasmas {\bf 22}, 013104 (2015).

\bibitem{roy05}
 P.~K.~Roy, S.~S.~Yu, E.~Henestroza, A.~Anders, F.~M.~Bieniosek, J.~Coleman, S.~Eylon, W.~G.~Greenway,
M.~Leitner, B.~G.~Logan, W.~L.~Waldron, D.~R.~Welch,
C.~Thoma, A.~B.~Sefkow, E.~P.~Gilson, P.~C.~Efthimion,
and R.~C.~Davidson, Phys. Rev. Lett. {\bf 95}, 234801 (2005).

\bibitem{startsev14}
 E.~A.~Startsev, I.~D.~Kaganovich, and R.~C.~Davidson,
Nucl. Instrum. Meth. A  {\bf 733}, 80 (2014).

\bibitem{sudan76}
 R.~N.~Sudan, Phys. Rev. Lett. {\bf 37}, 1613 (1976).
 
\bibitem{dorf09}
 M.~A.~Dorf, I.~D.~Kaganovich, E.~A.~Startsev, and R.~C.~Davidson, Phys. Rev. Lett. {\bf 103}, 075003 (2009).

\bibitem{mitrani14}
 J.~M.~Mitrani, I.~D.~Kaganovich, and R.~C.~Davidson,
Nucl. Instrum. Meth. A  {\bf 733}, 65 (2014).

\bibitem{tokluoglu15}
 E.~Tokluoglu and I.~D.~Kaganovich, Phys. Plasmas
{\bf 22}, 040701 (2015).

\bibitem{brunner14}
 S.~Brunner, R.~L.~Berger, B.~I.~Cohen, L.~Hausammann,
and E.~J.~Valeo, Phys. Plasmas {\bf  21}, 102104 (2014).

\bibitem{park15}
 J.~Park, D.~Caprioli, and A.~Spitkovsky, Phys. Rev. Lett.
{\bf 114}, 085003 (2015).

\bibitem{birdsall}
 C.~K.~Birdsall and A.~B.~Langdon, Plasma physics via
computer simulation (Institute of Physics, 2005).

\bibitem{hara15pop}
 K.~Hara, T.~Chapman, J.~W.~Banks, S.~Brunner,
I.~Joseph, R.~L.~Berger, and I.~D.~Boyd, Phys. Plasmas {\bf 22}, 022104 (2015).

\bibitem{hara17}
 K.~Hara, I.~Barth, E.~Kaminski, I.~Y.~Dodin, and N.~J.~Fisch, Phys.~Rev. E (accepted 2017).

\bibitem{buneman59}
 O.~Buneman, Phys. Rev. {\bf 115}, 503 (1959).

\bibitem{tokluoglu17}%
E.~Tokluoglu, I.~D.~Kaganovich, J.~Carlsson, K.~Hara,
and E.~A.~Startsev, Phys. Plasmas (submitted 2017).
\end{thebibliography}
\end{document}